\documentclass[12pt,thmsa]{article}

\input{tcilatex}
\begin{document}

\author{Paulo G. Macedo \\
{\small Centro de Astrofisica da U.P.}\\
{\scriptsize R. das Estrelas, s/n - 4150 Porto - PORTUGAL}}
\title{New Proposal for a 5-dimensional Unified Theory of Classical Fields of
Kaluza-Klein type}
\date{30/1/01}
\maketitle

\begin{abstract}
A new 5-dimensional Classical Unified Field Theory of Kaluza-Klein type is
formulated using 2 separate scalar fields which are related in such a way as
to make the 5-dimensional matter-geometry coupling parameter constant. It is
shown that this procedure solves the problem of the variability of the
gravity coupling parameter without having to assume a conformal invariance.

The corresponding Field equations are discussed paying particular attention
to the possible induction of scalar field gradients by Electromagnetic
Fields.

A new correspondence limit in which the field equations lead to the usual
Einstein-Maxwell equations is obtained. This limit does not require the
usual condition that the usual scalar field be constant.
\end{abstract}

\section{Introduction}

The aim of formulating a self consistent theory which unifies gravitation
and electromagnetism has been a long standing objective of Theoretical
Physics. In particular, Albert Einstein devoted a great deal of his life to
it.

It is our opinion that although we now know that there are other fields in
Nature apart from these two, namely the ones related with the strong and
weak interactions (which in the long run one aims to unify with these two if
one wants to formulate a complete Unified Field Theory), this is still a
useful exercise. In fact, as we shall show in later publications, this more
simple and completely classical construction still enables us to find
unexpected and interesting new experimental and observational predictions%
\footnote{%
We intend to analyze these problems in a later paper.}.

The first steps in trying to unify gravity with electromagnetism were given
by T. Kaluza\cite{Kaluza21} and O. Klein\cite{Klein26}. The theories
constructed by them describe the unified field in a 5-dimensional
Space-Time, whose metric, apart from the 4-dimensional metric, contains the
electromagnetic potential 4-vector. These theories contain as their field
equations the Einstein-Maxwell set and correctly describe the movement of
charged particles in the presence of Gravitational and Electromagnetic
fields.

However, although being mathematically very elegant, they were unable to
predict new effects which were unknown in the previous ones.

Later on, Jordan\cite{Jordan47}, Thiry\cite{Thiry48} and others like Bergmann%
\cite{Bergmann}, noticed that in order for the theory to be self-consistent,
a Scalar Field also had to come into play along with the Gravitational and
Electromagnetic fields. Their theory although not completely 5-dimensionally
covariant, was invariant under the action of the SO(3,1) and U(1) groups.
This smaller symmetry (more restricted than the SO(4,1) 5-dimensional
covariance) is due to the assumption of the existence of a Killing vector
field along the 5th coordinate (which is considered compact). This is
usually referred to as the cylindricity condition. Unlike the theory of
Kaluza and Klein these theories contain new information (in the form of a
15th field equation) unaccounted in the Einstein-Maxwell theory. In
particular they describe a complete coupling of the 3 (Gravitational ,
Electromagnetic and Scalar) fields (for a good review of the subject we
suggest the reading of the review article by Overduin and Wesson\cite{Wesson
97} and its bibliography).

These, more recent, compactified Kaluza-Klein type theories contain a scalar
field $\phi $ as well as a constant $\alpha $ which plays the role of making
the Electromagnetic potential 1-form $\mathbf{A}$ dimensionless in order to
be a part of the 5-dimensional metric. These theories contain 15 field
equations instead of the 14 contained in the original Kaluza-Klein
formulation.

In the limit when this Scalar field $\phi $ is constant, the first 14 field
equations reduce to the ones in Einstein-Maxwell theory. However, these
theories suffer from 2 main problems, which , in our opinion, haven 't yet
been solved in a satisfactory way, namely:

1) In the absence of a scalar field gradient the 15th field equation reduces
to the usual electromagnetic null field condition. This condition is
incompatible with the existence of electromagnetic fields in vacuum like the
pure electric field encountered in a planar capacitor, between the two
plates or the pure magnetic field such as the one in a solenoid. Therefore,
one is confronted with the problem that either one assumes $\phi $ to be
constant and recovers the Einstein-Maxwell theory, but is left with the
electromagnetic null field condition, or alternatively, one assumes that the
scalar field $\phi $ can vary and is faced with a second problem, namely:

2) In the 4-dimensional field equations which spring from the 5-dim
Einstein-Hilbert action, the geometry to matter source coupling $16\pi
G/c^{4}$ depends on the product $\alpha ^{2}\phi .$ Therefore, according to
these theories, this coupling should vary with the local scalar field value.
This raises the problem of the variability of Newton's Gravitational
coupling constant $G$. In fact, such variability has never been detected up
to now in our laboratory and Solar system experiments.

We should mention however that this does not necessarily mean that in very
large scales, outside the reach of our today's experimental range, $G$ or
even the speed of light $c,$ can not vary$.$ Indeed, to our opinion, there
has been a very wide-spread misunderstanding related to the constancy of the
speed of light $c.$ In fact this constancy, which springs from experiments
like the Michelson-Morley experiment is a well established fact, but the
great majority of authors has lost sight that such experiments are local
ones. In fact, this experiment only shows evidence that the in a given
Space-Time region the speed of light $c$ observed by two inertial observers
is the same irrespectively to their state of relative motion to each other.
To our knowledge, there is no evidence to acert that $c$ is a universal
constant (i.e. with the same value for every Space-Time region) and not a
local one. In fact, the speed of light $c$ is related to the electromagnetic
coupling, fine structure ''constant'' which recent results\cite{Barrow 98a}
seem to indicate to be not a constant at all but indeed vary with cosmic
time.

The usual way out to this $G$ variability problem consists in conformally
rescaling the metric in such a way as to make the $16\pi G/c^{4}$ coupling a
constant\cite{Wesson 97}.

\medskip In this paper, we suggest a different approach, in which the $%
\alpha $ parameter is not necessarily a constant but instead an auxiliary
scalar field which, for reasons that will become clear later on this paper,
we shall assume to be related with the $\phi $ field in such a way that $%
\alpha ^{2}\phi =16\pi G/c^{4}=const./R_{(5)}$ $,$ where $R_{(5)}$ is the
radius of the 5th dimension. The reason for this assumption has to do with
the fact that in a 5-dimensional gravity theory with 5-dim matter sources,
the coupling of the 5-dim Space-Time Geometry to its sources is given, as we
shall later see, by $\overline{\chi }=16\pi GR_{(5)}/c^{4}.$ The above
assumption only states that this coupling is a constant.

The assumption that the coupling parameter of the 5-dim Space-Time Geometry
to its sources $\overline{\chi }$ is a constant of Nature is very important
from the conceptual point of view. From the point of view of such a Unified
theory, it is perfectly conceivable that the parameters $G$ and $c$ which
characterize the couplings of the gravitational and electromagnetic
interactions separately could vary in different regions of Space-Time.
However, it seems to us much less reasonable to accept that the very
parameter $\overline{\chi }$ which describes the Unified Field coupling to
its sources could vary.

We shall further on in this paper show that this assumption leads to a
correspondence condition (occurring in the limit in which this theory is
equivalent to the Maxwell-Einstein Theory) which is the same as the one
obtain in the assumption that the 4-dimensional coupling parameter $\chi $
is constant. This is a different and much more general correspondence
condition then the one usually assumed, that $\phi =const.$.

In this way there is no need for a conformal rescaling of the metric.

In this approach, $\alpha $ is a parameter which measures the coupling
between the 5-dimensional Space-Time geometry and the Electromagnetic field
strength.

On the other hand, however, the dynamics of the scalar field has, to our
knowledge, received up till now a lesser degree of attention, particularly
in the presence of electromagnetic fields \cite{Macedo 90}$^{-}$\cite
{Mashhoon 96}.

Given the coupling between these fields which is clear in the field equations%
\cite{Wesson 97}, there exists the possibility of gradients of the scalar
field being induced by an electromagnetic field and therefore a scalar force
being generated\cite{Macedo 90}$^{-}$\cite{Mashhoon 96}. 

In the context of the theory that we propose in this paper, it is also
possible that, in specific circumstances, electromagnetic fields generate
local gradients of the scalar field. We shall show in a later publication
that this induction can result in several new effects like, in particular, a
slight modification of the dispersion relation of light waves propagating in
vacuum which would possibly explain the Ralston-Nodland claimed effect\cite
{Nodland98}.

Some of these effects have already been discussed by other authors%
\cite[ and ]{Mashhoon 96} and \cite{Gasperini87} in the context of usual
Kaluza-Klein theories.

This work, whose publication starts with this paper, shall be composed of
three papers:

1) In this first one we describe the proposed Classical Unified Field Theory
and discuss the correspondence limit in 5-dimensional vacuum.

2) In the second paper we shall concentrate specifically on wave solutions
and the propagation of electro-scalar waves in vacuum permeated by an
underlying uniform magnetic field looking in detail at the rotation of the
plane of polarization undergone by them as they propagate.

3) In the third paper, we shall try to formulate the Theory in the presence
of 5-dimensional Matter Sources and study the equation of motion of
particles subject to this Unified Field.

\medskip

\section{Formalism}

\subsection{The 5-dimensional metric}

\medskip

We shall start by assuming that a compactified point of view provides a good
description of our present day local region of 5-dimensional Space-Time.
I.e. the 5th dimension is therefore assumed to be topologically compact%
\footnote{%
This assumption does not in any way imply that its radius be very small. In
fact one can think of the 5th dimension as topologically compact but with a
characteristic curvature radius which can even be much larger than the
characteristic curvature radius of the 3 usual space dimensions.}.

As we' ve already mentioned, in some limit, we expect that this theory
should give us the Einstein - Maxwell Field Equations.. In order to achieve
that, following the usual procedure, we shall start by considering a
5-dimensional Space-Time whose metric $g_{_{AB}}$ and the corresponding
5-dimensional line element:

\begin{equation}
d\sigma ^{2}=\bar{g}_{_{AB}}\,dx^{A}\,dx^{B}  \label{5-d line element}
\end{equation}

\noindent (where $A,B,C,\dots =0,1,2,3,5$) and the metric $\bar{g}_{_{AB}}=%
\bar{g}_{_{AB}}(x^{C})$.

The extra dimension $x^{5}$ is assumed, as usual, to be space-like and
therefore we shall use a metric signature $(-\,,+\,,+\,,+\,,+)$.

From now on, and unless otherwise specified, we shall denote the
4-dimensional geometric object which corresponds to a given 5-dimensional
one by the same letter without a bar on top.

The 5-dimensional metric $\bar{g}_{_{AB}}$ must be such that from the field
tensors constructed out of it and its derivatives, one can obtain in a
specific limit the Einstein - Maxwell Field Equations.

In order to achieve this, it must be such that splitting it in the usual 4+1
way, the metric components can be written in the matrix form as follows:

\begin{equation}
\bar{g}_{_{AB}}=\left( 
\begin{array}{cc}
g_{\mu \nu }+{\alpha }^{2}\,\phi \,A_{\mu }\,A_{\nu }\  & \alpha \,\phi
\,A_{\mu } \\ 
\alpha \,\phi \,A_{\nu } & \phi
\end{array}
\right)  \label{metric 4+1}
\end{equation}

and its inverse can be written as:

\begin{equation}
\bar{g}^{^{AB}}=\left( 
\begin{array}{cc}
g^{\mu \nu } & -\,\alpha \,A^{\mu } \\ 
-\,\alpha \,A^{\nu } & {\alpha }^{2}\,A^{\beta }\,A_{\beta }\ +\frac{1}{\phi 
}
\end{array}
\right)  \label{inv metric 4+1}
\end{equation}

\medskip where the greek indices are 4-dimensional ones (i.e.$\mu ,\nu
,\dots =0,1,2,3)$ and where $A_{_{\mu }}$ is the 4-dimensional
electromagnetic 4-potential 1-form field,  $g_{_{\mu \nu }}$ is the usual
4-dimensional metric tensor field and $\phi $ and ${\alpha }$ are
4-dimensional scalar fields. The $\phi $ field is dimensionless, whereas the 
${\alpha }$ field needs to have the inverse dimensions of the
electromagnetic potential in order for the metric itself to be dimensionless.

\noindent i.e. 
\begin{eqnarray}
\phi &\equiv &{\bar{g}_{_{55}}}  \label{scalar field} \\
A_{_{\mu }} &\equiv &{\alpha }^{-1}\,\frac{{\bar{g}}_{_{\mu 5}}}{{\bar{g}}
_{_{55}}}  \label{E-M potential} \\
g_{_{\mu \nu }} &\equiv &{\bar{g}}_{_{\mu \nu }}-\frac{{\bar{g}}_{_{\mu 5}}\,%
{\bar{g}}_{_{\nu 5}}}{{\bar{g}}_{_{55}}}.  \label{4-d metric}
\end{eqnarray}

\medskip

\noindent In order for this 5-dimensional theory to be a viable one, one
would expect to find a specific limit where, in the case when $\bar{g}%
_{_{AB}}$ does not depend on $x^{5}$, (i. e. $\bar{g}_{_{AB,5}}\,=\,0,$
which characterizes compactified theories), the 5-dimensional vacuum
Einstein Equations  would lead to the Einstein-Maxwell's Equations for the $%
g_{\mu \nu }(x^{\rho })$ and $A_{\mu }(x^{\rho })$ fields.

From now on, we shall use instead of the $A_{_{\mu }}$ field another related
4-dimensional 1-form field $k_{_{\mu }}$ which we define as its dimensionless
counterpart:

\medskip

\begin{equation}
k_{\mu }=\alpha \,\ A_{\mu }
\end{equation}

\medskip

In this case, the 5-dimensional metric $\bar{g}_{_{AB}}$ is such that splits
in the usual 4+1 way:

\begin{equation}
\bar{g}_{_{AB}}=\left( 
\begin{array}{cc}
g_{\mu \nu }+\,\phi \,k_{\mu }\,k_{\nu }\  & \,\phi \,k_{\mu } \\ 
\,\phi \,k_{\nu } & \phi
\end{array}
\right)  \label{metric 4+1 with K}
\end{equation}

\medskip

its inverse being:

\begin{equation}
\bar{g}^{^{AB}}=\left( 
\begin{array}{cc}
g^{\mu \nu } & -\,k^{\mu } \\ 
-\,k^{\nu } & \,k^{\beta }\,k_{\beta }\ +\frac{1}{\phi }
\end{array}
\right)  \label{inv metric 4+1 with K}
\end{equation}

\noindent where$,$ therefore one can write: 
\begin{equation}
k_{\mu }\equiv \,\frac{\bar{g}_{\mu _{5}}}{\bar{g}_{_{55}}}.
\end{equation}

\subsection{The 5-dimensional Connection}

Starting from the above mentioned 5-dimensional metric, one can write the
corresponding 5-dimensional connection taking into account the metric and
its derivatives as follows: 
\begin{equation}
{\overline{\Gamma }}_{BC}^{A}=\frac{1}{2}\,\overline{g}^{AD}\,\left( 
\overline{g}_{_{BD,C}}+\overline{g}_{_{CD,B}}-\overline{g}_{_{BC,D}}\right)
\label{eq connexion}
\end{equation}

On the other hand, we shall also define a new 2-form field ${S_{\mu \beta }}$
which is the dimensionless correspondent of the Faraday field. It is obtained
form the dimensionless electromagnetic potential $k_{_{\mu }}$ as follows:

\begin{equation}
{S_{\mu \beta }}=\ {\ k_{\beta ,\mu }-k_{\mu ,\beta }}
\label{Adimensional Faraday tensor}
\end{equation}

i.e., this field is related with the Faraday field in the following way:

\begin{equation}
{S_{\mu \beta }}={\alpha \,F_{\mu \beta }+\alpha _{,\mu }\;A_{\,\beta
}-\alpha _{,\beta }}\ {{A}_{{\mu }}}
\end{equation}

\medskip

One can therefore write the 5-dimensional connection coefficients arising
from the above metric splitted as follows:

\begin{equation}
{\overline{\Gamma }}_{55}^{5}=\frac{1}{2}\,k^{\mu }\,\phi _{,\mu }
\label{connection555}
\end{equation}

\begin{equation}
{\overline{\Gamma }}_{55}^{\mu }=-\,\frac{1}{2}\,\phi ^{,\mu }
\label{eq connexionmu55b}
\end{equation}

\begin{equation}
\overline{\Gamma }_{\beta 5}^{\mu }\;=\;{\frac{1}{2}}\,\left( {\phi
\,S_{\beta }^{\;\,\mu }\;-\;\phi ^{\,,\mu }\,k_{\,\beta }}\right) {\ }
\label{connectionmubeta5}
\end{equation}

\begin{equation}
\overline{\Gamma }_{\beta 5}^{5}\;=\;\,{\frac{1}{2}}\left[ {\phi \,k^{\mu
}\,S_{\mu \beta }\;+\;{\frac{{\phi _{,\beta }}}{\phi }}\;+\;\phi _{,\mu
}\,k^{\mu }\,k_{\,\beta }{\ }}\right]  \label{connection5beta5}
\end{equation}

\begin{equation}
\overline{\Gamma }_{\beta \sigma }^{\mu }=\Gamma _{\beta \sigma }^{\mu }+%
\frac{\phi }{2}\left( k_{\beta }\ S_{\sigma }^{\hspace{4pt}\mu }+k_{\sigma
}\ S_{\beta }^{\hspace{4pt}\mu }\right) -\frac{1}{2}\phi ^{,\mu }k_{\sigma
}k_{\beta }  \label{connectionmubetasigma}
\end{equation}

and

\begin{equation}
{\overline{\Gamma }}_{\beta \sigma }^{5}=\frac{1}{2}\left\{ 
\begin{array}{c}
\frac{1}{\phi }\,\left[ \,\phi _{,\sigma }\,k_{\beta }+\,\phi _{,\beta
}\,k_{\sigma }\right] +\,k^{\mu }\,\phi _{,\mu }\,k_{\beta }\,k_{\sigma }\,
\\ 
+\phi \ k^{\mu }\,\,\,\,\left[ S_{\mu \sigma }\,k_{\beta }+S_{\mu \beta
}\,k_{\sigma }\right] +\,\,k_{\beta ;\sigma }+\,k_{\sigma ;\beta }
\end{array}
\right\} \,  \label{connection5betasigma}
\end{equation}

\medskip

\subsection{The 5-dimensional \protect\medskip Ricci Tensor.}

From the above connection and its derivatives, one can construct the
5-dimensional Ricci Tensor:

\begin{equation}
\overline{R}_{BD}\equiv \overline{R}^{A}\,_{BAD}\equiv {\overline{\Gamma }}
_{BD,A}^{A}-{\overline{\Gamma }}_{BA,D}^{A}+{\ \ \overline{\Gamma }}%
_{BD}^{E}\,{\overline{\Gamma }}_{AE}^{A}-{\overline{\Gamma }}_{BA}^{E}\,{%
\overline{\Gamma }}_{DE}^{A}  \label{eq Riemann}
\end{equation}

Whose components can be splitted as follows: 
\begin{equation}
\overline{R}_{55}=-\,\frac{1}{2}\,\phi _{\ \ \ \ ;\rho }^{,\rho }{\,{\ }+\ 
\frac{1}{4}\,{\frac{{\phi ^{,\rho }\phi _{,\rho }}}{\phi }}}+\,\,{\,\frac{{%
\phi }^{2}}{4}}\,{\,S_{\rho \sigma }^{\;\,}\,S^{\rho \sigma }}
\label{eq Ricci55}
\end{equation}

\begin{equation}
\overline{R}_{5\mu }=\,{\frac{3}{4}\ {\phi }_{{,\rho }}\ \,S_{\mu
}^{\;\,\rho }+\frac{1}{2}\ \phi \,S_{\mu \ \ ;\ \rho }^{\;\,\rho }}+k_{\mu
}\ \overline{R}_{55}  \label{eq Ricci5mu}
\end{equation}

\begin{eqnarray}
\overline{R}_{\mu \nu } &=&R_{\mu \nu }-{\frac{1}{2}}\left( {{\frac{{\phi
_{;\mu \nu }}}{\phi }}\;-\frac{1}{2}\ }\frac{{\phi _{,\mu }\phi _{,\nu }}}{
\phi ^{2}}\right) -{\frac{1}{2}}\,{\phi \,S_{\mu }^{\;\,\rho }S_{\nu \rho }}
\nonumber \\
&&+\overline{R}_{5\mu }k_{\mu }\,+\overline{R}_{5\nu }k_{\nu }\ -\overline{R}%
_{55}{\ }k_{\mu }k_{\nu }  \label{eq Riccimunu}
\end{eqnarray}

\medskip

\subsection{5-Dimensional Vacuum Field Equations}

\medskip We shall now look at 5-Dimensional Vacuum Field Equations and see
what new information they contain. In particular we shall look at the
coupling of the three (Gravitational, Electromagnetic and Scalar) fields.

One can now write the 5-dimensional Einstein's Field Equations in vacuum as:

\begin{equation}
\overline{R}_{BD}\;=\;0  \label{eq 5d vacuum}
\end{equation}

Using the components of the 5-dimensional Ricci tensor given by equations (%
\ref{eq Ricci55}) to (\ref{eq Riccimunu}), one can write the field equations
(\ref{eq 5d vacuum}) as follows:

\medskip

\begin{equation}
-\,\,\phi _{\ \ \ \ ;\rho }^{,\rho }{\,{\ }+\ \frac{1}{2}\,{\frac{{\phi
^{,\rho }\phi _{,\rho }}}{\phi }}}+\,\,{\,\frac{{\phi }^{2}}{2}}\,{\,S_{\rho
\sigma }^{\;\,}\,S^{\rho \sigma }}=0  \label{eq Ricci55=0}
\end{equation}

\begin{equation}
\,{\frac{3}{2}\ {\phi }_{{,\rho }}\ \,S_{\mu }^{\;\,\rho }+\ \phi \,S_{\mu \
\ ;\ \rho }^{\;\,\rho }}=0  \label{eq Ricci5mu=0}
\end{equation}

and

\begin{equation}
R_{\mu \nu }-{\frac{1}{2}}\left( {{\frac{{\phi _{;\mu \nu }}}{\phi }}\;-%
\frac{1}{2}\ }\frac{{\phi _{,\mu }\phi _{,\nu }}}{\phi ^{2}}\right) -{\frac{1%
}{2}}\,{\phi \,S_{\mu }^{\;\,\rho }S_{\nu \rho }}=0  \label{eq Riccimunu=0}
\end{equation}

\medskip

\medskip We shall now try to find a correspondence limit where the last two
sets of field equations (\ref{eq Ricci5mu=0}) and (\ref{eq Riccimunu=0})
give us the usual Maxwell - Einstein set of field equations.

Substituting the dimensionless electromagnetic tensor ${S_{\nu \rho },}$ by
its expression given in (\ref{Adimensional Faraday tensor}), in the above
Einstein type Equations (\ref{eq Riccimunu=0}) one obtains:

\medskip 
\begin{eqnarray}
R_{\mu \nu } &=&{\frac{1}{2}}\left( {{\frac{{\phi _{;\mu \nu }}}{\phi }}\;-%
\frac{1}{2}\ }\frac{{\phi _{,\mu }\phi _{,\nu }}}{\phi ^{2}}\right) 
\nonumber \\
&&+{\frac{1}{2}}\,{\phi \,}\left[ 
\begin{array}{c}
{\alpha }^{2}{\,F_{\mu }^{\;\,\rho }F_{\nu \rho }+\alpha \,F_{\mu
}^{\;\,\rho }\left( {\alpha _{,\nu }\;A_{\,\rho }-\alpha _{,\rho }}\ {{A}
_{\nu }}\right) } \\ 
{+}\left( {\alpha _{,\mu }\;A_{\,}^{\rho }-\alpha _{\,}^{,\rho }}\ {{A}_{{\
\mu }}}\right) \left( {\alpha \,F_{\nu \rho }+\alpha _{,\nu }\;A_{\,\rho
}-\alpha _{,\rho }}\ {{A}_{\nu }}\right)
\end{array}
\right]  \label{Generalized Einstein}
\end{eqnarray}

\medskip

\medskip on the other hand, substituting (\ref{Adimensional Faraday tensor})
in the above Maxwell's type equations (\ref{eq Ricci5mu=0}), one obtains 
\begin{eqnarray}
0 &=&\,{\phi \,\alpha \,F_{\mu \ \ ;\ \rho }^{\;\,\rho }+}\left( {\frac{3}{2}%
\ {\phi }_{{,\rho }}\ \alpha +\ \phi \,{\alpha }_{,\rho }}\right) {\,F_{\mu
}^{\;\,\rho }}  \nonumber \\
&&{+\,\,{\frac{3}{2}\ {\phi }_{{,\rho }}}\left( {\alpha _{,\mu
}\;A_{\,}^{\rho }-\alpha _{\,}^{,\rho }}\ {{A}_{{\mu }}}\right) +\ \phi
\left( {\alpha _{,\mu }\;A_{\,}^{\rho }-\alpha _{\,}^{,\rho }}\ {{A}_{{\mu }}%
}\right) }_{;\ \rho }  \label{Generalized Maxwell}
\end{eqnarray}

\medskip

\medskip In order for equations (\ref{Generalized Einstein}) and (\ref
{Generalized Maxwell}) to give us the set of the usual Maxwell - Einstein
field equations, one must have the 3 following correspondence conditions:

\begin{eqnarray}
&&\,\left( {\frac{3}{2}\ {\phi }_{{,\rho }}\ \alpha +\ \phi \,{\alpha }
_{,\rho }}\right) {\,F_{\mu }^{\;\,\rho }}  \nonumber \\
&&{+\,\,{\frac{3}{2}\ {\phi }_{{,\rho }}}\left( {\alpha _{,\mu
}\;A_{\,}^{\rho }-\alpha _{\,}^{,\rho }}\ {{A}_{{\mu }}}\right) +\ \phi
\left( {\alpha _{,\mu }\;A_{\,}^{\rho }-\alpha _{\,}^{,\rho }}\ {{A}_{{\mu }}%
}\right) }_{;\ \rho }=0,  \label{Correspondence 1}
\end{eqnarray}

\begin{equation}
{\alpha \,F_{\mu }^{\;\,\rho }\left( {\alpha _{,\nu }\;A_{\,\rho }-\alpha
_{,\rho }}\ {{A}_{\nu }}\right) +}\left( {\alpha _{,\mu }\;A_{\,}^{\rho
}-\alpha _{\,}^{,\rho }}\ {{A}_{{\mu }}}\right) \left( {\alpha \,F_{\nu \rho
}+\alpha _{,\nu }\;A_{\,\rho }-\alpha _{,\rho }}\ {{A}_{\nu }}\right) =0
\label{Correspondence 2}
\end{equation}

and

\begin{equation}
{\frac{{\alpha }^{2}{\phi }}{2}}\,{\,\,=\ }\chi =\frac{8\pi G}{c^{4}}
\label{Correspondence 3}
\end{equation}

\medskip

the first two above conditions (\ref{Correspondence 1}) and (\ref
{Correspondence 2}) can be substituted by the following :

\begin{equation}
\,\left( {\frac{3}{2}\ {\phi }_{{,\rho }}\ \alpha +\ \phi \,{\alpha }_{,\rho
}}\right) {\,F_{\mu }^{\;\,\rho }}=0  \label{Correspondence 1a}
\end{equation}

and

\begin{equation}
{{\alpha _{,\nu }\;A_{\,\rho }-\alpha _{,\rho }}\ {{A}_{\nu }}}=0
\label{Correspondence 2a}
\end{equation}

One can rewrite this condition by splitting it in its space and time
components as follows:

\begin{equation}
\left\{ 
\begin{array}{c}
{\varepsilon }^{ijk}{{\ \alpha _{,j}}\ {{A}_{k}}}=0 \\ 
{{\alpha _{,0}\;A_{\,i}=\alpha _{,i}}\ {{A}_{0}}}
\end{array}
\right.  \label{Correspondence 2b}
\end{equation}

where $\mathbf{\varepsilon }$ is the Space 3-dimensional Levi-Civita tensor.
This means that, in this classic correspondence limit, the spatial part of
the $\alpha $ scalar field gradient is alined with the electromagnetic
potential $\mathbf{A}$\textbf{.}

\medskip On the other hand, looking at the 5-dim. metric, one can see that
the scalar field $\phi $ is such that $\sqrt{\phi }$ plays the role of the
5th dimension scale factor$.$ Therefore, the 5th dimension radius $R_{(5)}$,
is related to the scalar field $\phi $ in the following way:

\begin{equation}
R_{(5)}\propto \sqrt{\phi }  \label{5th dimension radius}
\end{equation}

Since in this type of 5-dimensional Kaluza-Klein theories the 5-dimensional
and 4-dimensional gravity coupling parameters ${\chi }$ and $\overline{\chi }
$ are related in such a way that:

\begin{equation}
{\overline{\chi }=\chi \ }R_{(5)}  \label{Chi-Chi bar relation}
\end{equation}

Under the above assumption (that ${\overline{\chi }}$ is constant), by
substituting (\ref{Correspondence 3}) and (\ref{5th dimension radius}) on (%
\ref{Chi-Chi bar relation}), one can write:

\begin{equation}
\alpha ^{2}\phi ^{\frac{3}{2}}=const.  \label{chi-bar condition}
\end{equation}

\medskip Since both $\alpha \ $and ${\phi }$ fields are always non null, one
can differentiate this equation and divide both members by $\alpha \ \phi ^{%
\frac{1}{2}},$ obtaining:

\begin{equation}
\,{\frac{3}{2}\ \ \alpha \ {\phi }_{{,\rho }}+\ 2\,{\alpha }_{,\rho }\phi }=0
\label{Chi bar condition 2}
\end{equation}

One can therefore substitute (\ref{Correspondence 1a}) in the above equation
(\ref{Chi bar condition 2}) , and divide by $\alpha $ , obtaining that:

\medskip 
\begin{equation}
{{\phi }_{{,\rho }}\ \,F_{\mu }^{\;\,\rho }}=0
\label{Phi-F correspondence condition}
\end{equation}

\medskip

Analyzing this above correspondence condition, one can see that:

For $\mu =0,$ one obtains:

\begin{equation}
\nabla \phi \ \bullet\ \mathbf{E=0}
\end{equation}

This means that, in this correspondence limit, the gradient of the scalar
field $\phi $ is orthogonal to the Electric field.

On the other hand, for $\mu =i,$ one obtains:

\medskip

\begin{equation}
{\phi }^{{,0}}{\ \mathbf{E=}\,c}\nabla \phi \ \times\ \mathbf{B}
\end{equation}

This means that, in this correspondence limit and in the absence of an
electric field, or, in the case when the scalar field $\phi $ is stationary,
the gradient of the scalar field $\phi $ is alined with the magnetic field,
and conversely, in the absence of either a magnetic field or a scalar field
spatial gradient, either the electric field is null or the scalar field is
stationary.

In fact, substituting (\ref{Correspondence 1}) in (\ref{Generalized Maxwell}
) and dividing both members by ${\alpha \ \phi ,}$ one can also see that the
set of field equations, in this correspondence limit, can be written as
follows:

\begin{equation}
F_{\mu }^{\,\beta }{}_{;\beta }\;=\;0  \label{limit eq Ricci5mu=0}
\end{equation}

\begin{equation}
R_{\mu \nu }={\frac{1}{2}}\,{\frac{{\phi _{;\mu \nu }}}{\phi }}\;-\,\,{\frac{%
1}{4}}\,\,{\frac{{\phi _{,\mu }\,\phi _{,\nu }}}{{\phi ^{2}}}}\;-\,{\frac{{\
\alpha ^{2}\phi }}{2}}\,F_{\mu }^{\,\beta }\,F_{\nu \beta }
\label{limit eq Riccimunu=0}
\end{equation}

\begin{equation}
2\,\;\frac{\phi _{;\beta }^{\;\;\beta }}{\phi }-\,{\frac{{\phi _{,\beta
}\,\phi ^{,\beta }}}{{\phi ^{2}}}}\;\,=\alpha ^{2}\phi \,F_{\mu \beta
}\,F^{\,\mu \beta }\;\;  \label{limit eq Ricci55=0}
\end{equation}

and the correspondence condition is given by equation (\ref{Phi-F
correspondence condition}).

\medskip I.e., one can see that in this limit, one obtains the last 14
equations as the Einstein-Maxwell field Equations. for vacuum permeated by
an electromagnetic field, provided that the identification given by equation
(\ref{Correspondence 3}) is carried out, where $\alpha \,$and $\phi $ are
the local space-time values of the corresponding $\alpha \,$and $\phi $
scalar fields.

One can also see that in this correspondence limit, the 15th field equation
takes the form:

\begin{equation}
\,\,\frac{\phi _{\ \ \ \ ;\rho }^{,\rho }}{\phi }{\,{\ -}\ \frac{1}{2}\,{%
\frac{{\phi ^{,\rho }\phi _{,\rho }}}{{\phi }^{2}}}}=\,\,{\,\,}\frac{8\pi G}{
c^{4}}\,{\,F^{^{\rho \,\sigma }}\,F_{\sigma \,\rho }^{\;}}  \label{15th eq.}
\end{equation}

which can also be written in terms of the electric and magnetic
3-dimensional vector fields in the following way:

\begin{equation}
\frac{\phi _{\ \ \ \ ;\rho }^{,\rho }}{\phi }{\,{\ -}\ \frac{1}{2}\,{\frac{{%
\ \phi ^{,\rho }\phi _{,\rho }}}{{\phi }^{2}}}}\,\,{\,}={\,}\frac{8\pi G}{
c^{4}}\,{\,}\left( \mathbf{E}^{2}-c^{2}\mathbf{B}^{2}\right)
\label{15th eq.a}
\end{equation}

\medskip As an exercise, one can see that if instead one would assume, as
usual, that $\chi $ is itself constant (instead of $\overline{\chi }),$ one
would have, the following condition:

\begin{equation}
\alpha ^{2}\phi =const.  \label{chi condition}
\end{equation}

\medskip instead of the condition given by equation. (\ref{chi-bar condition}
). Therefore, differentiating this equation  and dividing by $\alpha $ , 
instead of equation (\ref{Chi bar condition 2}), one would obtain:

\begin{equation}
{\ \ \alpha \ {\phi }_{{,\rho }}+\ 2\,{\alpha }_{,\rho }\phi }=0
\end{equation}

substituting this result in the above condition (\ref{Correspondence 1a}),
one would again obtain the same correspondence conditions given by (\ref
{Correspondence 2a}) and (\ref{Phi-F correspondence condition}).

I.e., in both cases (namely, taking $\chi =const.$ or $\overline{\chi }%
=const.)$ one obtains the same correspondence conditions. These conditions
are in no way restricted to the usually assumed condition that $\phi =const.$%
.

One can see that, in this theory, since in the classical correspondence
limit one can still have a varying $\phi $ field, one is therefore no longer
bound by the electromagnetic null field condition ($F_{\mu \beta }\,F^{\,\mu
\beta }\;=\;0$) which in the usual formulations arise from the 15th field
equation (\ref{15th eq.}).

This electromagnetic null field condition ($\mathbf{E}^{2}-c^{2}\mathbf{B}%
^{2}=0$) which is present in other Kaluza-Klein formulations is obviously
too restrictive , since one knows that it is possible to generate in vacuum,
pure stationary electric or magnetic uniform fields like the ones present in
a plane plate condenser or inside a solenoid. These electromagnetic fields
do not satisfy the above mentioned null field condition.

One can therefore see that, according to this theory, even uniform and
stationary electromagnetic fields must induce local scalar field gradients.

In the next publication of this series of 3 papers, we shall explicitly
describe the solutions of the field equations corresponding to these two
simple cases.

\medskip

\section{Conclusions}

To summarize, we conclude that:

In this proposed theory, the parameter $\alpha $ measures the coupling
between the 5-dimensional Space-Time geometry and the Electromagnetic field
strength and plays the role of rendering its product with the
Electromagnetic potential 1-form $\mathbf{A}$ dimensionless in order to be a
part of the 5-dimensional metric.  By assuming $\alpha $  to be an auxiliary
scalar field instead of assuming it to be a constant, we were able to avoid
the usual need for conformal invariance of the 5-dim. Kaluza -Klein type
Theories. As we have recalled, this need arised from the fact that it was
thought that in the correspondence limit in which the 5-dimensional gravity
theory would yield the usual Einstein- Maxwell theory, the $\phi $ scalar
field would have to be constant. This in turn led to the fact that, in this
limit, the 15th field equation implied the electromagnetic null field
condition $c^{2}B^{2}-E^{2}=0$. This condition, for the above mentioned
reasons, was shown to be too restrictive.

In this proposal, this condition is substituted by a much less restrictive
correspondence condition, namely ${{\phi }_{{,\rho }}\ \,F_{\mu }^{\;\,\rho }%
}=0.$ This allows for this 5-dimensional gravity theory, in the
correspondence limit, to describe such simple electromagnetic fields as the
ones arising in solenoids or between the plates of a plane plate condenser.
These were not possible to account for in the usual Kaluza-Klein type
formulation.

\medskip The subject is however far from being thoroughly studied. In
particular, one can ask what is the physical meaning and the physical
reasons leading to this new correspondence condition ${{\phi }_{{,\rho }}\
\,F_{\mu }^{\;\,\rho }}=0$. We would expect this condition to arise
naturally from the initial assumption of the cilindricity condition (which
amounts to the assumption of the existence of a Killing vector field of the
5-dimensional metric). This could be connected to the arisal of an
anisotropy in 5-dimensional Space-Time in the Early Universe. This would in
turn generate the splitting of the 4 space dimensions in 3+1 that we observe
today.

This problem along with the analysis of the Equation of motion for particles
subject to the electromagnetic, scalar and gravitational fields fields, as
well as the analysis of wave solutions for the theory and the formulation of
the theory in the presence of 5-dimensional matter sources will be the
subject of a series of publications of which this is the first one.

\medskip

\section{Acknowledgements}

\medskip The author would like to acknowledge the financial support of
''Funda\c{c}\~{a}o para a Ciencia e Tecnologia'' through the
PESO/P/PRO/1195/97 research project.

The research work whose publication is started by this first paper has been
a long standing one, and a special word of thanks goes to my former students
Carlos Herdeiro, Carlos Martins and Nuno Barreiros as well as my collegues
Caroline S. Silva, Paulo Crawford do Nascimento and Eduardo Lage for their
help in revising my calculations and for useful discussions.

\bigskip


\medskip

\begin{thebibliography}{99}
\bibitem{Kaluza21}  T. Kaluza, Zum Unit\"{a}tsproblem der Physik , Sitz.
Preuss. Akad. Wiss. Phys. Math. K1, 966, (1921).

\bibitem{Klein26}  O. Klein,Quantentheorie und f\"{u}nfdimensionale
Relativit\"{a}tstheorie , Zeits. Phys. , 37, 895, (1926)

\bibitem{Jordan47}  P. Jordan, Erweiterung der Projektiven
Relativit\"{a}tstheorie , Ann.. Phys. (Leipzig) , 1, 219, (1947).

\bibitem{Thiry48}  Y. Thiry, Les \'{e}quations de la Th\'{e}orie Unitaire de
Kaluza , Comptes Rendus Acad. Sci. (Paris) , 226, 216, (1948)

\bibitem{Bergmann}  P. Bergmann, Unified Field Theory with fifteen variables
,Ann. Maths., 49, 255, (1948).

\bibitem{Wesson 97}  J.M. Overduin and P.S. Wesson, Kaluza-Klein Gravity,
Phys. Reports, 283, 303-378, (1997).

\bibitem{Macedo 90}  P. G. Macedo, Electric Neutrality and the Jordan Thiry
Scalar Field., Phys. Lett. A. Vol. 143, nb. 3, 96, (1990).

\bibitem{Macedo 91}  P. G. Macedo, Bounds for Jordan-Thiry Scalar Field
Coupling Constant, Phys. Lett. A. Vol. 156, nb. 1, 2 , 20, (1991).

\bibitem{Wesson 95}  P.S. Wesson and J. Ponce de Leon, The equations of
motion in Kaluza-Klein cosmology and its implications for astrophysics, A.
\& A., 294 (1995), 1.

\bibitem{Mashhoon 96}  B. Mashhoon, P.S. Wesson and H. Liu, Fifth force from
fifth dimension, Univ. of Missouri preprint (1996).

\bibitem{Gasperini87}  M. Gasperini, Experimental Tests on Unified theories
of the scalar-vector-tensor type , Phys. Rev. D 36, nb. 8, 2318 (1987) 3202.

\bibitem{Nodland98}  Borge Nodland and John P. Ralston, Indication of
Anisotropy in Electromagnetic Propagation over Cosmological Distances ,
Phys. Rev. Lett. 78, 3043, (1997).

\bibitem{Barrow 98a}  J. K. Webb et All., Evidence for Time Variation of the
Fine Structure Constant, Astro-ph/9803165.
\end{thebibliography}
\end{document}